\documentclass[aps,preprint,showkeys]{revtex4}%
\usepackage{amsfonts}
\usepackage{amsmath}
\usepackage{amssymb}
\usepackage{graphicx}
\usepackage{float}
\usepackage{epstopdf}%
\providecommand{\U}[1]{\protect\rule{.1in}{.1in}}
\begin{document}
\title[ ]{Singular inverse square potential in coordinate space\\with a minimal length}
\author{Djamil Bouaziz}
\affiliation{Department of Physics, University of Jijel, BP 98, Ouled Aissa, 18000 Jijel, Algeria}
\author{Tolga Birkandan}
\affiliation{Department of Physics, Istanbul Technical University, 34469 Istanbul, Turkey}
\pacs{PACS number}

\begin{abstract}
The problem of a particle of mass $m$ in the field of the inverse square
potential $\alpha/r^{2}$ is studied in quantum mechanics with a generalized
uncertainty principle, characterized by the existence of a minimal length.
Using the coordinate representation, for a specific form of the generalized
uncertainty relation, we solve the deformed Schr\"{o}dinger equation
analytically in terms of confluent Heun functions. We explicitly show the
regularizing effect of the minimal length on the singularity of the potential.
We discuss the problem of bound states in detail and we derive an expression
for the energy spectrum in a natural way from the square integrability
condition; the results are in complete agreement with the literature.

\end{abstract}
\volumeyear{2017}
\volumenumber{number}
\issuenumber{number}
\eid{identifier}
\startpage{1}
\endpage{ }
\maketitle







\section{Introduction}

In spite of its singular features, the inverse square potential $\alpha/r^{2}$
remains one of the most important interactions in quantum mechanics. This
potential function appears in the study of many important problems in
different fields of physics, such as Efimov effect \cite{1.1}, dipole-bound
anions in polar molecules \cite{1.2,1.22}, atoms interacting with a charged
wire \cite{1.3}, the analysis of the near-horizon structure of black holes
\cite{12} and the interaction of a dipole in a cosmic string background
\cite{djamil2}.

The singularity of this potential manifests in the solutions of the
Schr\"{o}dinger equation where the square integrability condition does not
lead to an orthogonal set of eigenfunctions with their corresponding
eigenenergies \cite{case}. This is because the Hamiltonian operator for a
singular $\alpha/r^{2}$ potential is not essentially self-adjoint \cite{metz},
and therefore, to restore a discrete bound states spectrum, one must apply the
method of self-adjoint extensions or equivalently require orthogonality of the
wave functions \cite{case}. Besides, alternative approaches have been used to
deal with this potential, such as the standard regularization methods
\cite{1.22,reg,camblong} or the renormalization techniques \cite{gub,beane}.
Furthermore, it has been shown in Refs. \cite{djamil2,djamil1} that this
potential becomes regular in the framework of quantum mechanics with a
generalized uncertainty principle (GUP) \cite{magiore,k1,k2,k3,k4} due to the
presence of a minimal length, which plays the role of a regularizing cutoff at
short distances. In this context, the deformed Schr\"{o}dinger equation has
been solved in momentum space and the energy spectrum was computed in a
natural way \cite{djamil1}.

It is noteworthy that the generalization of the Heisenberg uncertainty
principle in order to include an elementary length is a common prediction of
various studies in quantum gravity \cite{alden,garay,rov} and string theory
\cite{21,amati,konishi}. The implications of such modifications on the
mathematics of quantum mechanics have first been addressed by Kempf and his
collaborators in several papers \cite{k1,k2,k3,k4}. Since then, many studies
have been directed to investigate the theoretical and the physical
consequences of the GUP, see, for instance, Refs. \cite{quesne,ali,bouk,phys}.
Particularly, special attention was given to the fundamental non-relativistic
quantum systems, such as the harmonic oscillator \cite{k1,brau,chang}, the
hydrogen atom in one \cite{fit,p} and three \cite{brau,sandor,stetsko,boua3}
dimensions, the singular inverse square potential \cite{djamil2,djamil1} and
the gravitational quantum well \cite{brau2,N}. In most of these works, the
Schr\"{o}dinger equation has been solved in momentum space as the coordinate
representation leads to a fourth order differential equation, whose resolution
is generally not possible. This explains the use of the perturbation
techniques in the study of certain problems in coordinate space
\cite{bouk,brau,stetsko,brau2,N}. However, there are some disagreements
between the results of the coordinate and momentum representations, as that
observed in the case of the hydrogen atom \cite{boua3}. Let us note that the
perturbative approach cannot be applied for the $\alpha/r^{2}$ potential to
investigate the minimal length corrections in coordinate space. Therefore this
potential has been studied only in momentum space. On another side, it has
been recently proposed, in Ref. \cite{ha}, an ad hoc transformation to reduce
the order of the Schr\"{o}dinger differential equation in coordinate space.
Then, this approach has been applied to the spherical square well potential to
investigate the consequence of the GUP on the resonant scattering.

In this paper, we use the transformation of Ref. \cite{ha} to study the
inverse square potential $\alpha/r^{2}$ in coordinate space in the presence of
a minimal length. We show that the deformed Schr\"{o}dinger equation can be
solved analytically in terms of confluent Heun functions which are also found
as solutions in ordinary quantum mechanical systems \cite{ronv,slav,hort,ishk}%
. Then we illustrate the regularizing effect of the minimal length and we
investigate the problem of bound states; the energy spectrum will be computed
by simply requiring a physical boundary condition. Before doing so, let us
mention that this study allows us to check whether the momentum and coordinate
representations lead to the same results in the case of the $\alpha/r^{2}$ potential.

In Sec. II, we review the basic mathematics of quantum mechanics with a GUP.
Sec. III is devoted to the study of the deformed Schr\"{o}dinger equation for
the $\alpha/r^{2}$ potential in coordinate space where the effect of the
minimal length on the singularity in this problem will be examined in detail.
In Sec. IV, we investigate the problem of bound states and derive an
expression of the energy spectrum. In the last section, we summarize our
results and conclusions.

\section{Schr\"{o}dinger equation in coordinate space with a GUP}

Diverse forms of the generalized uncertainty principle (GUP) have been
proposed in the literature: there is a GUP with a minimal length \cite{k1}, a
GUP which incorporates a minimal length and a minimal momentum \cite{k3}, a
GUP with a Lorentz-covariant algebra \cite{quesne}, and a GUP including a
minimal length and a maximal momentum \cite{ali}. In this work, we are
interested in the first form, where the GUP can be expressed in $N$-dimensions
as \cite{djamil1}:
\begin{equation}
\left(  \Delta X_{i}\right)  \left(  \Delta P_{i}\right)  \geq\frac{\hbar}%
{2}\left(  1+\beta%
{\textstyle\sum\limits_{j=1}^{N}}
[\left(  \Delta P_{j}\right)  ^{2}+\left\langle \widehat{P}_{j}\right\rangle
^{2}]+\beta^{^{\prime}}[\left(  \Delta P_{i}\right)  ^{2}+\left\langle
\widehat{P}_{i}\right\rangle ^{2}]\right)  ,\label{gu}%
\end{equation}
where $\beta$ and $\beta^{\prime}$ are small positive parameters. This GUP can
be obtained from the following modified Heisenberg algebra
\cite{k1,k2,chang,djamil1}:
\begin{align}
\lbrack\widehat{X}_{i},\widehat{P}_{j}] &  =i\hbar\lbrack(1+\beta\widehat
{P}^{2})\delta_{ij}+\beta^{^{\prime}}\widehat{P}_{i}\widehat{P}_{j}],\text{
\ \ \ }[\widehat{P}_{i},\widehat{P}_{j}]=0,\nonumber\\
\lbrack\widehat{X}_{i},\widehat{X}_{j}] &  =i\hbar\frac{2\beta-\beta
^{^{\prime}}+\beta(2\beta+\beta^{^{\prime}})\widehat{P}^{2}}{1+\beta
\widehat{P}^{2}}(\widehat{P}_{i}\widehat{X}_{j}-\widehat{P}_{j}\widehat{X}%
_{i}).\label{1}%
\end{align}
The GUP (\ref{gu}) implies the existence of a minimal length, given by
\cite{k2}%
\begin{equation}
\left(  \Delta X_{i}\right)  _{\min}=\hbar\sqrt{N\beta+\beta^{^{\prime}}%
},\text{ \ \ }\forall i.\label{3}%
\end{equation}
One of the most used representations of the position and momentum operators
satisfying the commutation relations (\ref{1}) are \cite{k2,chang}%
\begin{equation}
\widehat{X}_{i}=i\hbar\lbrack\left(  1+\beta p^{2}\right)  \frac{\partial
}{\partial p_{i}}+\beta^{\prime}p_{i}p_{j}\frac{\partial}{\partial p_{j}%
}+\gamma p_{i}],\text{ \ \ \ \ \ }\widehat{P}_{i}=p_{i},\label{7}%
\end{equation}
where $\gamma$ is a small positive parameter related with $\beta$ and
$\beta^{\prime}$.

As mentioned in Sec. I, the inverse square potential was studied in Refs.
\cite{djamil1,djamil2} by the help of representation (\ref{7}).

In coordinate space, up to the first order in $\beta,$ the operators
$\widehat{X}_{i}$ and $\widehat{P}_{i}$ can be represented by \cite{stetsko}%
\begin{equation}
\widehat{X}_{i}=\widehat{x}_{i}+\frac{2\beta-\beta^{\prime}}{4}(\widehat
{p}^{2}\widehat{x}_{i}+\widehat{x}_{i}\widehat{p}^{2}),\text{ \ }\widehat
{P}_{i}=\widehat{p}_{i}(1+\frac{\beta^{\prime}}{2}\widehat{p}^{2}), \label{cr}%
\end{equation}
where $\widehat{x}_{i}$ and $\widehat{p}_{i}$ satisfy the standard commutation
relations of ordinary quantum mechanics.

In the case $\beta^{\prime}=2\beta,$ representation (\ref{cr}) reduces to
\begin{equation}
\widehat{X}_{i}=\widehat{x}_{i},\text{ \ \ \ }\widehat{P}_{i}=\widehat{p}%
_{i}\left(  1+\beta\widehat{p}^{2}\right)  , \label{brau}%
\end{equation}
which was firstly used in Ref. \cite{brau} to study the hydrogen atom. In this
special case, the deformed algebra (\ref{1}) takes the form%
\begin{equation}
\lbrack\widehat{X}_{i},\widehat{P}_{j}]=i\hbar\lbrack(1+\beta\widehat{P}%
^{2})\delta_{ij}+2\beta\widehat{P}_{i}\widehat{P}_{j}],\text{ }[\widehat
{P}_{i},\widehat{P}_{j}]=0,\text{ }[\widehat{X}_{i},\widehat{X}_{j}%
]=0,\nonumber
\end{equation}
and the minimal length reads in 3-dimensions as
\begin{equation}
\left(  \Delta X_{i}\right)  _{\min}=\hbar\sqrt{5\beta},\text{ \ \ }\forall i.
\end{equation}
Let us now write the Schr\"{o}dinger equation for a particle of mass $m$ in
the external potential $V(r)$ in coordinate space by using the representation
(\ref{brau}) as follows
\begin{equation}
\lbrack\frac{\widehat{p}^{2}}{2m}+V(r)+\frac{\beta}{m}\widehat{p}^{4}%
]\psi(\overset{\rightarrow}{r})=E\psi(\overset{\rightarrow}{r}), \label{6}%
\end{equation}
where the terms of order $\beta^{2}$ have been neglected.

Equation (\ref{6}) is a fourth order differential equation, and thereby its
resolution is not obvious in general. However, it has been recently shown in
Ref. \cite{ha} that the order of Eq. (\ref{6}) can be reduced by performing
the following transformation:%
\begin{equation}
\psi(\overset{\rightarrow}{r})=(1-2\beta\widehat{p}^{2})\phi(\overset
{\rightarrow}{r}).
\end{equation}
The function $\phi(\overset{\rightarrow}{r})$ then satisfies the equation
\begin{equation}
\left[  \left(  1+4m\beta\lbrack E-V(r)]\right)  \frac{\widehat{p}^{2}}%
{2m}+V(r)-E\right]  \phi(\overset{\rightarrow}{r})=0.
\end{equation}
By using the factorization $\phi(\overset{\rightarrow}{r})=R(r)Y_{\ell}%
^{m}(\theta,\varphi)$, the radial function $R(r)$\ satisfies the second order
differential equation
\begin{equation}
\left[  \left(  1+4m\beta\lbrack E-V(r)]\right)  \left(  \frac{d^{2}}{dr^{2}%
}+\frac{2}{r}\frac{d}{dr}-\frac{\ell(\ell+1)}{r^{2}}\right)  +\frac{2m}%
{\hbar^{2}}[E-V(r)]\right]  R(r)=0. \label{ds}%
\end{equation}
In the following, we will use this equation to study the $\alpha/r^{2}$
potential in coordinate space with a minimal length depending on the
deformation parameter $\beta$.

\section{Inverse square potential in coordinate space with a GUP}

The deformed Schr\"{o}dinger equation (\ref{ds}) for a particle of mass $m$ in
the field of the attractive inverse square potential $V(r)=-\frac{\alpha
}{r^{2}},$ ($\alpha>0$), reads%
\begin{equation}
\left[  \left(  1-\Omega-\frac{\alpha\Omega}{Er^{2}}\right)  \frac{d^{2}%
}{dr^{2}}+\frac{2}{r}\left(  1-\Omega-\frac{\alpha\Omega}{Er^{2}}\right)
\frac{d}{dr}-\left(  1-\Omega-\frac{\alpha\Omega}{Er^{2}}\right)  \frac
{L}{r^{2}}+\frac{2mE}{\hbar^{2}}\left(  \frac{\alpha}{Er^{2}}+1\right)
\right]  R(r)=0,\label{dse}%
\end{equation}
with definitions $L=\ell(\ell+1)$ and $\Omega=-4m\beta E.$

Let us examine the effect of the deformation parameter $\beta$ on the
asymptotic behaviors of the two linearly independent solutions of Eq.
(\ref{dse}) in the regions $r\approx0$ and $r\rightarrow\infty.$

In the region $r\approx0,$ we write Eq. (\ref{dse}) by keeping only the
dominant terms as
\begin{equation}
\lbrack\frac{d^{2}}{dr^{2}}+\frac{2}{r}\frac{d}{dr}-\frac{L}{r^{2}}]R(r)=0.
\label{ce}%
\end{equation}
By putting $R(r)=r^{s}$ in Eq. (\ref{ce}), we get two values of $s$
\ ($s=-\frac{1}{2}\pm\frac{1}{2}\sqrt{1+4L}$), corresponding to two solutions%
\begin{equation}
R_{1}(r\ll1)=r^{-1-\ell},\ R_{2}(r\ll1)=r^{\ell}. \label{ab}%
\end{equation}
When $\ell\neq0$, the function $R_{1}$ is not square integrable in the origin.
In the case where $\ell=0$, the two solutions are square integrable and the
situation is analogous to that discussed by Landau and Lifshitz \cite{landau},
where it was concluded that one must always choose the solution which is less
divergent at the origin. Therefore, the physical solution has to be chosen as
$R_{2}$.

Note that these behaviors are completely different from that of ordinary
quantum mechanics, where there is a difference between two ranges of the
coupling $\alpha$ of the potential: the weak coupling regime $2m\alpha
/\hbar^{2}<1/4+L$ and the strong coupling regime $2m\alpha/\hbar^{2}\geq
1/4+L$. In the second regime the two solutions behave similarly at the origin
and, both of them are quadratically integrable. Now, in the deformed case the
second solution $R_{2}$ is the physical one regardless the value of the
coupling constant $\alpha$. This might be viewed as an indication of the
regularization of the singular inverse square potential in this formalism of
the GUP.

In the region $r\rightarrow\infty,$ we proceed in the same manner. We write
Eq. (\ref{dse}) by keeping only the dominant terms as
\begin{equation}
\lbrack\frac{d^{2}}{dr^{2}}+\frac{2}{r}\frac{d}{dr}+\frac{2mE}{\hbar
^{2}\left(  1-\Omega\right)  }]R(r)=0. \label{ei}%
\end{equation}
We use the transformation $R(r)=\frac{1}{r}u(r)$, then $u(r)$ satisfies the
equation
\begin{equation}
\lbrack\frac{d^{2}}{dr^{2}}+\frac{2mE}{\hbar^{2}\left(  1-\Omega\right)
}]u(r)=0.
\end{equation}
For bound states ($E<0$), the solutions of this equation are clearly given by
\begin{equation}
u(r)=\exp\bigg(\pm\sqrt{-\frac{2mE}{\hbar^{2}\left(  1-\Omega\right)  }%
}r\bigg). \label{abi}%
\end{equation}
These asymptotic behaviors are analogous to that of the solutions of ordinary
quantum mechanics ($\beta=0$): one of the solutions behaves at infinity as
$u_{1}(r)\sim\exp(-\eta r),$ with $\eta^{2}=-\frac{2mE}{\hbar^{2}}$, and it is
square integrable, and the other solution has the behavior $u_{2}(r)\sim
\exp(\eta r),$ which is not physically acceptable. This result is expected
since the minimal length modifies the characteristics of the potential only at
short distances, and it has no effect at large distances.

We return now to Eq. (\ref{dse}). By introducing the dimensionless variable
$x=-\frac{Er^{2}}{\alpha}$, and with the definitions $\kappa=\frac{m\alpha
}{2\hbar^{2}},$ $\varepsilon=1-\Omega,$ equation (\ref{dse}) takes the form
\begin{equation}
\left[  \left(  \varepsilon x+\Omega\right)  \frac{d^{2}}{dx^{2}}+\frac{3}%
{2}\frac{(\varepsilon x+\Omega)}{x}\frac{d}{dx}-\frac{L}{4}\frac{(\varepsilon
x+\Omega)}{x^{2}}+\kappa\frac{(1-x)}{x}\right]  R(x)=0. \label{eq}%
\end{equation}
We perform the transformation
\begin{equation}
R(x)=x^{-\frac{1}{4}\sqrt{4L+1}-\frac{1}{4}}f(x)=x^{-\frac{1}{2}\ell-\frac
{1}{2}}f(x), \label{trr}%
\end{equation}
and the equation (\ref{eq}) becomes
\begin{equation}
\left[  \frac{d^{2}}{dx^{2}}+\frac{1/2-\ell}{x}\frac{d}{dx}+\frac
{\kappa\left(  1-x\right)  \allowbreak}{x\left(  \varepsilon x+\Omega\right)
}\right]  f(x)=0.
\end{equation}
To move the singular point $-\frac{\Omega}{\varepsilon}$ to the value $1,$ we
simply change our variable as
\begin{equation}
y=-\frac{\varepsilon}{\Omega}x,
\end{equation}
which leads to the equation%
\begin{equation}
\left(  \frac{d^{2}}{dy^{2}}+\frac{1/2-\ell}{y}\frac{d}{dy}+\frac{\frac
{\kappa}{\varepsilon}\left(  \frac{\Omega}{\varepsilon}y+1\right)
\allowbreak}{y\left(  y-1\right)  }\right)  f(y)=0.
\end{equation}
We then apply another transformation
\begin{equation}
f(y)=\left(  y-1\right)  g(y), \label{t2}%
\end{equation}
and the equation for $g(y)$ is
\begin{equation}
\left(  \frac{d^{2}}{dy^{2}}+\left[  \frac{1/2-\ell}{y}+\frac{2}{\left(
y-1\right)  }\right]  \frac{d}{dy}+\frac{\frac{\kappa}{\varepsilon}\left(
\frac{\Omega}{\varepsilon}y+1\right)  +1/2-\ell}{y\left(  y-1\right)
}\right)  g(y)=0. \label{he}%
\end{equation}

\bigskip Equation (\ref{he}) is in the form of the following confluent Heun
equation \cite{ronv,slav}
\begin{equation}
\left[  \frac{d^{2}}{dy^{2}}+\left(  a+\frac{b+1}{y}+\frac{c+1}{\left(
y-1\right)  }\right)  \frac{d}{dy}+\frac{\left(  \frac{1}{2}a(b+c+2)+d\right)
y+e+\frac{b}{2}+\frac{1}{2}\left(  c-a\right)  (b+1)}{y\left(  y-1\right)
}\right]  g(y)=0, \label{h}%
\end{equation}
with the parameters
\begin{align}
a  &  =0,\text{ \ \ }b=-1/2-\ell,\nonumber\\
c  &  =1,\text{ \ \ }d=\frac{\kappa\Omega}{\varepsilon^{2}},\text{
\ \ }e=\frac{\kappa}{\varepsilon}+1/2.
\end{align}

Equation (\ref{h}) is a second order linear differential equation with two
regular singularities at $y=0$ and $1$, and an irregular singularity of rank
$1$ at $y=\infty$. In the vicinity of $y=0$, the two linearly independent
solutions of Eq. (\ref{h}) are \cite{ronv}
\begin{equation}
g_{1}(y)=Hc(a,b,c,d,e;y),\text{ \ \ }g_{2}(y)=y^{-b}Hc(a,-b,c,d,e;y),
\end{equation}
where $Hc$ is the confluent Heun function. According to the transformations
(\ref{trr}) and (\ref{t2}), one has
\begin{equation}
R(y)=y^{-\frac{1}{2}\ell-\frac{1}{2}}(y-1)g(y), \label{tfi}%
\end{equation}
and consequently the deformed Schr\"{o}dinger equation (\ref{dse}) admits two
solutions as
\begin{align}
R_{1}(y)  &  =y^{-\frac{\ell}{2}-\frac{1}{2}}(y-1)Hc(a,b,c,d,e;y),\label{s1}\\
R_{2}(y)  &  =y^{\frac{\ell}{2}}(y-1)Hc(a,-b,c,d,e;y), \label{s2}%
\end{align}
where%
\begin{equation}
y=-\frac{1+4m\beta E}{4m\beta\alpha}r^{2}%
\end{equation}

In the vicinity of $r=0$ ($y=0$), the confluent Heun function behaves as
$Hc(a,b,c,d,e;0)=1.$ So the asymptotic behaviors of the two solutions in this
region are
\begin{equation}
R_{1}(y)\approx y^{-\frac{\ell}{2}-\frac{1}{2}}(y-1)\approx r^{-\ell-1},\text{
\ }R_{2}(y)\approx y^{\frac{\ell}{2}}(y-1)\approx r^{\ell},
\end{equation}
which is exactly what we have already obtained in Eq. (\ref{ab}). It was
indicated that $R_{1}(y)$ is not square integrable, so that the physical
solution is $R_{2}(y)$. Then the solution of the deformed Schr\"{o}dinger
equation (\ref{dse}) reads
\begin{equation}
R(y)=y^{\frac{\ell}{2}}(y-1)Hc(a,-b,c,d,e;y),
\end{equation}
or in terms of the old variable $r$ as%
\begin{equation}
R(r)=Ar^{\ell}\bigg(1+\frac{1+4m\beta E}{4m\beta\alpha}r^{2}%
\bigg)Hc\bigg(a,-b,c,d,e;-\frac{1+4m\beta E}{4m\beta\alpha}r^{2}\bigg),
\label{fs}%
\end{equation}
where $A$ is a normalization constant, and the parameters are given by%
\begin{align}
a  &  =0,\text{ }b=-1/2-\ell,\text{ }c=1,\text{ }d=\frac{\kappa\Omega
}{\varepsilon^{2}}=-\frac{4m^{2}\alpha\beta E}{2\hbar^{2}(1+4m\beta E)^{2}%
},\nonumber\\
\text{ \ \ }e  &  =\frac{\kappa}{\varepsilon}+1/2=\frac{m\alpha}{2\hbar
^{2}(1+4m\beta E)}+1/2,\nonumber
\end{align}
in their explicit form.

As illustrated above, the potential is now regularized by the introduction of
the minimal length parameter $\beta$. Accordingly, the quantized energy
spectrum should be the manifestation of a physical boundary condition, as it
will be shown in the following section.

\section{Energy of bound states}

To obtain the energy spectrum we should guarantee the square integrability
condition on the whole interval of $r$ for all values of eigenenergies $E$. We
must, however, take into account that the solution (\ref{fs}) is well-defined
only for $\left\vert y\right\vert <1$ as it should be for a local Frobenius
solution \cite{slav}. Then the asymptotic form of the wave-function in the
region ($r\rightarrow\infty$) cannot be examined from the solution (\ref{fs})
as $\infty$ is an irregular singularity of the confluent Heun equation. Around
the irregular singular point at $\infty$, one can define Thom{\'{e}}-type
(asymptotic series) solutions, which have the form given in Eq. (\ref{abi}).
To overcome this difficulty, one needs to define a range of the radial
distance $r$ compatible with the physical constraints for studying the
asymptotic form of the solution within this range.

To this end, we can consider, as in Refs. \cite{beane}, very low energy levels
such as $\sqrt{\frac{-2mE}{\hbar^{2}}}R\ll1$ (here the cutoff $R$ is the
minimal length $\left(  \Delta x\right)  _{\min}=\hbar\sqrt{5\beta}$), which
implies that $\Omega$ $\ll1$.

Let us now rewrite the function (\ref{fs}) as
\begin{equation}
R(r)=Ar^{\ell}\bigg(1-\frac{(\Omega-1)r^{2}}{4m\beta\alpha}%
\bigg)Hc\bigg(a,-b,c,d,e;\frac{(\Omega-1)r^{2}}{4m\beta\alpha}\bigg),
\label{fsa}%
\end{equation}
where $\Omega=-4m\beta E$ with $\Omega>0$ for bound states. It can be observed
that for infinite values of $r$, the confluent Heun series diverges because
the argument of the function, in Eq. (\ref{fsa}) grows up, and therefore, a
physical condition should be associated with the solution (\ref{fsa}). To deal
with this problem, let us focus on large values of $r$ such as $\frac{r^{2}%
}{4m\beta\alpha}\rightarrow\frac{1}{\Omega}\gg1$ (or $r\rightarrow\sqrt
{\frac{-\alpha}{E}}$), and require the constraint $R(r=\sqrt{\frac{-\alpha}%
{E}})=0$, which yields the following spectral condition,
\begin{equation}
Hc\bigg(a,-b,c,d,e;\frac{\Omega-1}{\Omega}\bigg)=0, \label{sc}%
\end{equation}
which gives the bound state energy levels of the $\alpha/r^{2}$ potential in
the presence of a minimal length.

It would be important to emphasize that a mathematical reasoning lead to the
condition (\ref{sc}). Recalling that the wave-function (\ref{fsa}) corresponds
to a series solution of the confluent Heun equation around $y=0$, where the
radius of convergence is found to be $\left\vert y\right\vert =1$ \cite{ronv}.
Therefore, in the limit $r\rightarrow c\sqrt{\frac{-\alpha}{E}}$ ($c$ being a
positive constant), one has $y\rightarrow c\frac{\Omega-1}{\Omega}$, and so if
$\Omega<\frac{c}{c+1}$ then $\left\vert y\right\vert >1$, which implies that
the confluent Heun series diverges. In our considerations, $c=1$ and
$\Omega\ll1,$ consequently, the series always diverges; hence one needs to
impose the condition (\ref{sc}).

Before examining Eq. (\ref{sc}), let us show that in this limit (i.e.,
$\Omega\ll1$), the confluent Heun function can be reduced to a hypergeometric
function by doing the following approximations:
\begin{equation}
d=\frac{\kappa\Omega}{(1-\Omega)^{2}}\simeq0,\text{ \ }e=\frac{\kappa
}{(1-\Omega)}+1/2\simeq\kappa+1/2,
\end{equation}
In this case the confluent Heun equation (\ref{h}) reduces to
\begin{equation}
\left[  \frac{d^{2}}{dy^{2}}+\left(  \frac{b+1}{y}+\frac{c+1}{y-1}\right)
\frac{d}{dy}+\frac{e+\frac{b}{2}+\frac{c}{2}(b+1)}{y\left(  y-1\right)
}\right]  g(y)=0,
\end{equation}
which is a hypergeometric equation of the form \cite{abramo}%
\begin{equation}
\left[  y\left(  1-y\right)  \frac{d^{2}}{dy^{2}}+\bigg(\delta-(\alpha
+\gamma+1)y\bigg)\frac{d}{dy}-\alpha\gamma\right]  g(y)=0,
\end{equation}
where the parameters $\alpha$, $\gamma$ and $\delta$ are
\begin{equation}
\alpha=\frac{3}{4}-\frac{\ell}{2}-i\frac{\nu}{2},\text{ \ }\gamma=\frac{3}%
{4}-\frac{\ell}{2}+i\frac{\nu}{2}\text{, \ }\delta=\frac{1}{2}-\ell,
\end{equation}
with $\nu=$ $\sqrt{4\kappa-\left(  \ell+\frac{1}{2}\right)  ^{2}}$ and
$\kappa=\frac{m\alpha}{2\hbar^{2}}.$ By using the transformation (\ref{tfi}),
the two solutions now take the forms \cite{abramo}
\begin{align}
R_{1}(y)  &  =y^{-\frac{\ell}{2}-\frac{1}{2}}(y-1)F(\alpha,\gamma,\delta;y),\\
R_{2}(y)  &  =y^{-\frac{\ell}{2}-\frac{1}{2}}(y-1)y^{1-\delta}F(\alpha
^{\prime},\gamma^{\prime},\delta^{\prime};y),
\end{align}
where
\begin{align*}
\alpha^{\prime}  &  =\alpha-\delta=\frac{1}{4}\allowbreak+\frac{\ell}{2}%
-\frac{i\nu}{2},\\
\gamma^{\prime}  &  =\gamma-\delta=\frac{1}{4}\allowbreak+\frac{\ell}{2}%
+\frac{i\nu}{2},\\
\delta^{\prime}  &  =2-\delta=\frac{3}{2}+\ell.
\end{align*}

It follows that the regular solution at the origin is $R_{2}(y)$, which can be
written as follows%
\begin{equation}
R(y)\big{|}_{\Omega\ll1}=A_{0}y^{\frac{\ell}{2}}(y-1)F\bigg(\alpha^{\prime
},\gamma^{\prime},\delta^{\prime};y\bigg), \label{wa}%
\end{equation}
where $A_{0}$ is a normalization constant.

\bigskip\ Now, by using the expression (\ref{wa}), the associated boundary
condition that substitutes for Eq. (\ref{sc}) is then
\begin{equation}
F\bigg(\alpha^{\prime},\gamma^{\prime},\delta^{\prime};-\frac{1}{\Omega
}\bigg)\underset{\Omega\ll1}{=}0, \label{qcc}%
\end{equation}
By using the following transformation \cite{abramo}\
\begin{align}
F(\alpha,\gamma,\delta;z)=  &  \frac{\Gamma(\delta)\Gamma(\gamma-\alpha
)}{\Gamma(\gamma)\Gamma(\delta-\alpha)}(-z)^{-\alpha}F\bigg(\alpha
,1-\delta+\alpha,1-\gamma+\alpha;\frac{1}{z}\bigg)\nonumber\\
&  +\frac{\Gamma(\delta)\Gamma(\alpha-\gamma)}{\Gamma(\alpha)\Gamma
(\delta-\gamma)}(-z)^{-\gamma}F\bigg(\gamma,1-\delta+\gamma,1-\alpha
+\gamma;\frac{1}{z}\bigg),
\end{align}
and by taking $F(a,b,c;y)\underset{y\ll1}{\approx}1$, the equation (\ref{qcc})
can be rewritten in the form%
\begin{equation}
\left\vert B\right\vert \left(  \frac{1}{\Omega}\right)  ^{-\frac{1}%
{4}\allowbreak-\frac{\ell}{2}}\Gamma\bigg(\frac{3}{2}+\ell\bigg)\left[  \exp
i\left(  \left\{  \frac{\nu}{2}\ln(\frac{1}{\Omega})+\arg(B)\right\}  \right)
+\exp\left(  -i\left\{  \frac{\nu}{2}\ln(\frac{1}{\Omega})+\arg(B)\right\}
\right)  \right]  =0, \label{14}%
\end{equation}
where
\begin{equation}
B\equiv\frac{\Gamma(i\nu)}{\Gamma(\frac{1}{4}\allowbreak+\frac{\ell}{2}%
+\frac{i\nu}{2})\Gamma(\frac{5}{4}\allowbreak+\frac{\ell}{2}+\frac{i\nu}{2}%
)}=\left\vert B\right\vert \exp[i\arg(B)].
\end{equation}
Then, Eq. (\ref{14}) leads to the condition
\begin{equation}
\cos\bigg[\arg(B)+\frac{\nu}{2}\ln(\frac{1}{\Omega})\bigg]=0, \label{15}%
\end{equation}
which gives the following energy spectrum:%
\begin{equation}
E_{n}=-\frac{1}{4m\beta}\exp\frac{2}{\nu}\left[  \arg(B)-(n+\frac{1}{2}%
)\pi\right]  ,\text{ \ }n=0,1,2,...., \label{es}%
\end{equation}
which can be written in terms of the minimal length $\left(  \Delta x\right)
_{\min}=\hbar\sqrt{5\beta}$ as%
\begin{equation}
E_{n}=-\frac{5\hbar^{2}}{4m\left(  \Delta x\right)  _{\min}^{2}}\exp\frac
{2}{\nu}\left[  \arg(B)-(n+\frac{1}{2})\pi\right]  ,\text{ \ }n=0,1,2,.....
\label{ess}%
\end{equation}

The expression (\ref{es}) is identical to the one obtained in momentum space
in Ref. \cite{djamil1} by using the representation (\ref{7}). This result is
also reached in standard quantum mechanics by regularization techniques, where
the potential is cut off at a short-distance radius $R$, and the potential is
replaced in the region $r<R$ by another interaction, see, for instance, Refs.
\cite{1.22,reg,gub}. It follows that the minimal length plays the same role as
the ultraviolet cutoff $R$. In the standard regularization methods, the limit
$R\rightarrow0$ does not make sense because the energy goes to $-\infty$ and
this explains the need to perform renormalization \cite{reg,gub}. However, in
the minimal length formalism, the limit\ $\beta\rightarrow0$ is not mandatory
as $\beta$ is a physical parameter of this formalism, and hence it should
appear in expressions of the observable quantities.

As it has been pointed out in Ref. \cite{djamil1}, the condition $\left\vert
E_{n}\right\vert \ll1/4m\beta$ systematically excludes the undesirable values
of the number $n$ in the formula (\ref{es}), so that there is now a ground
state with finite energy. Furthermore, if the potential is weakly attractive
($4\kappa<1/4$) there exists no bound state solution for Eq. (\ref{qcc}).

To confirm these results in coordinate space, we will now examine the exact
spectral equation (\ref{sc}). We have plotted the confluent Heun function in
Eq. (\ref{sc}) as a function of $\omega=\Omega/2=-2m\beta E$ for different
values of the parameter $\kappa=m\alpha/2\hbar^{2}$, by taking $\ell=0$ as in
Ref. \cite{djamil1}. The zeros of the function represent the eigenenergies of
the potential. For the sake of comparison with the momentum space results, we
choose the same values of $\kappa$ as in Ref. \cite{djamil1}, namely,
$\kappa=1/20,$ $3/4$ and $2.$ One can claim that the energy of the ground
state is finite and, as in ordinary quantum mechanics, there are many, almost
identical, excited states with $\Omega\simeq0$ (accumulation point). Indeed,
in Fig. 1, corresponding to $\kappa=3/4$, the energy of the ground state is
proportional to $\omega_{1}\approx\allowbreak0.0491$ (in Ref. \cite{djamil1},
one has $\omega_{1}\approx0.0694$); and for $\kappa=2$, Fig. 2 shows that
$\omega_{1}\approx0.2486$ (we have $\omega_{1}\approx0.3704$ in Ref.
\cite{djamil1}).

This difference between the results is expected as the two spectral
conditions, namely the Eq. (\ref{sc}) and that of Ref. \cite{djamil1}
correspond to the two special cases $\beta^{\prime}=2\beta$ and $\beta
^{\prime}=\beta$ respectively; and in the general case ($\beta^{\prime}%
\neq\beta$), the spectrum inversely depends on the sum $\beta^{\prime}+\beta$
\cite{djamil1}, giving rise to the factor $3/2$ between the two results. On
another side, the momentum representation used in Ref. \cite{djamil1} is exact
while the one used here satisfies the GUP algebra in the first order of the
deformation parameter.

For $\kappa=1/20$, Fig. 3 shows that there is no bound state; the value of the
critical coupling constant, below which bound states do not exist, is also
identical to what was obtained in Ref. \cite{djamil1} ($\kappa^{\ast}=1/16$
for $\ell=0$), which is the same as in ordinary quantum mechanics.

\begin{figure}[ptb]
\centering
\includegraphics[scale=0.4]{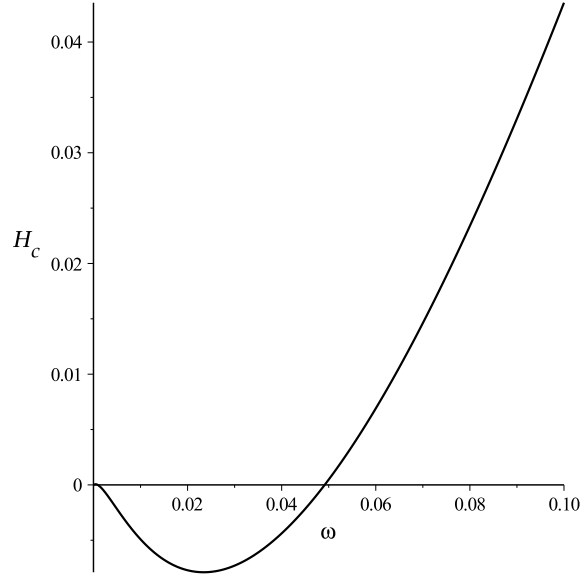} \caption{$H_{C}\equiv
Hc\bigg(a,-b,c,d,e;\frac{2\omega-1}{2\omega}\bigg)$ for $\kappa=3/4;$
$\omega=-2m\beta E$.}%
\label{fig:fig1}%
\end{figure}

\begin{figure}[ptb]
\centering
\includegraphics[scale=0.4]{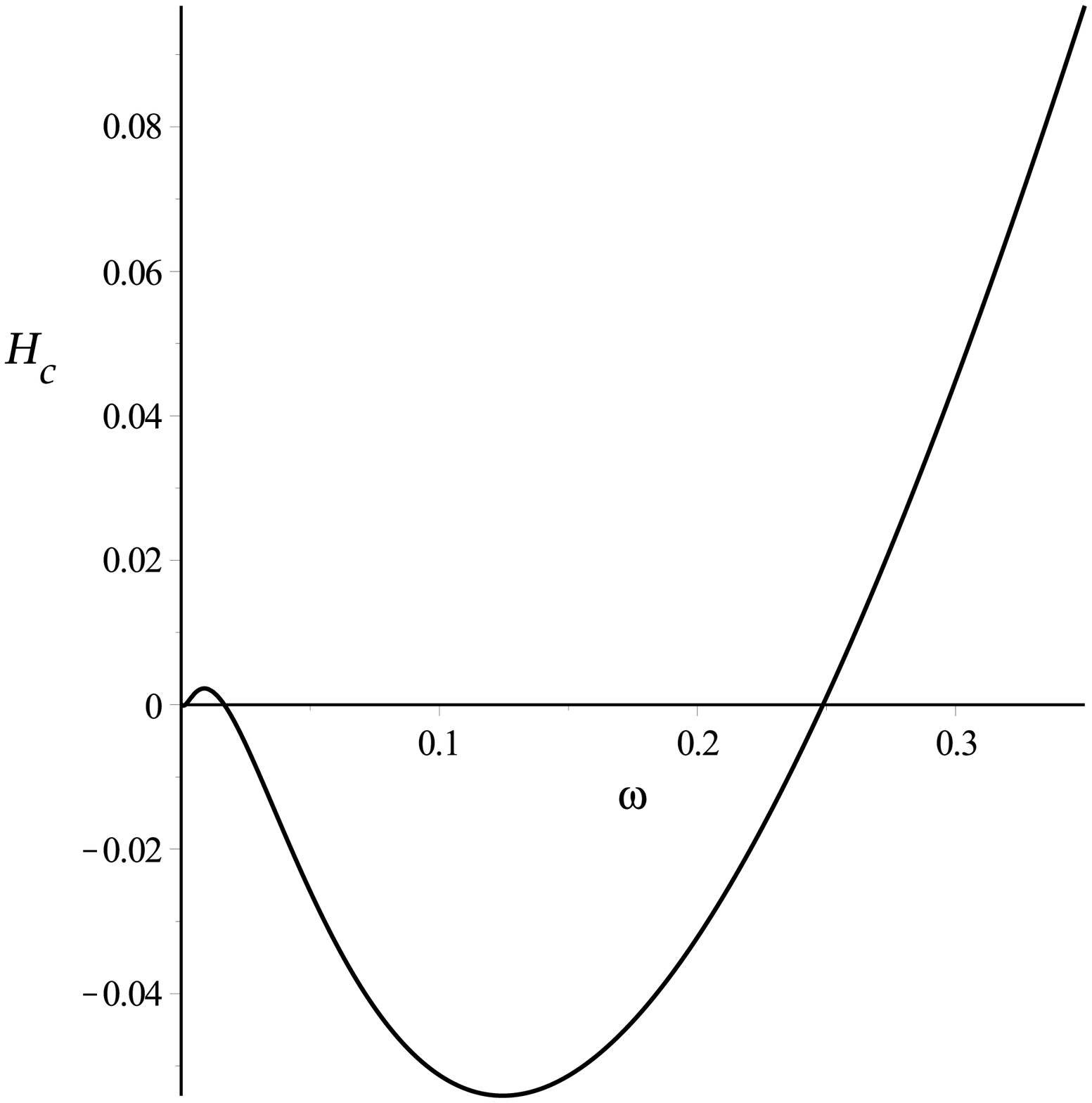} \caption{$H_{C}\equiv
Hc\bigg(a,-b,c,d,e;\frac{2\omega-1}{2\omega}\bigg)$ for $\kappa=2;$
$\omega=-2m\beta E$.}%
\label{fig:fig2}%
\end{figure}

\begin{figure}[ptb]
\centering
\includegraphics[scale=0.4]{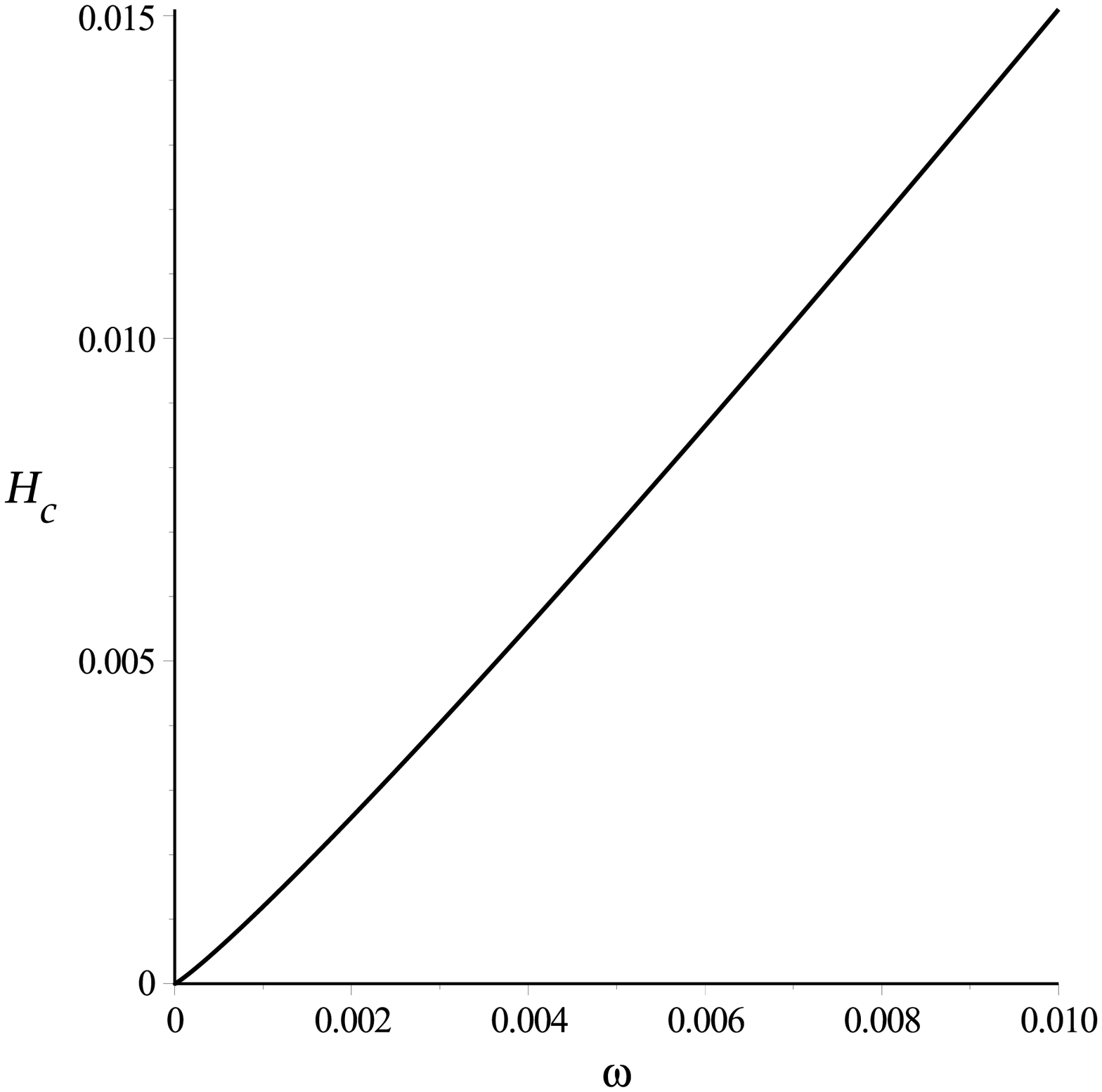} \caption{$H_{C}\equiv
Hc\bigg(a,-b,c,d,e;\frac{2\omega-1}{2\omega}\bigg)$ for $\kappa=1/20;$
$\omega=-2m\beta E$.}%
\label{fig:fig3}%
\end{figure}

To complete this analysis, it is important to present some curves of the
radial wave-function in coordinate space to show in particular that the
boundary condition (\ref{sc}) leads to a decaying behavior for large values of
$r$ such as $r\rightarrow\sqrt{\frac{-\alpha}{E}}$. We plotted the radial
wave-function in Eq. (\ref{fsa}) as a function of the dimensionless variable
$\xi=\frac{r}{\left(  \Delta x\right)  _{\min}}$ with $\kappa=2$, for
different values of the parameter $\omega=-2m\beta E$ (solutions and not
solutions of the boundary condition (\ref{sc}) by taking $\ell=0$). Thus, for
the two eigenvalues $\omega=0.0167$ and $0.000167$, Figs. 4 and 5 show,
respectively, the decaying behavior of $R(\xi)$ when $\xi$ grows up. In
contrast, Fig. 6 with the value $\omega=0.004,$ which is not a solution of the
quantization condition (\ref{sc}), shows that $R(\xi)$ does not decay for
large values of $\xi$ in the aforementioned range of the radial variable.

This study shows that the coordinate space representation of Ref. \cite{brau},
used firstly to study the hydrogen atom, gives the same results as that
obtained in Ref. \cite{djamil1} in the case of the inverse square potential by
using the momentum representation. This result is intriguing because these two
representations led to conflicting results in some problems \cite{boua3}, such
as the hydrogen atom case. \begin{figure}[ptb]
\centering
\includegraphics[scale=0.4]{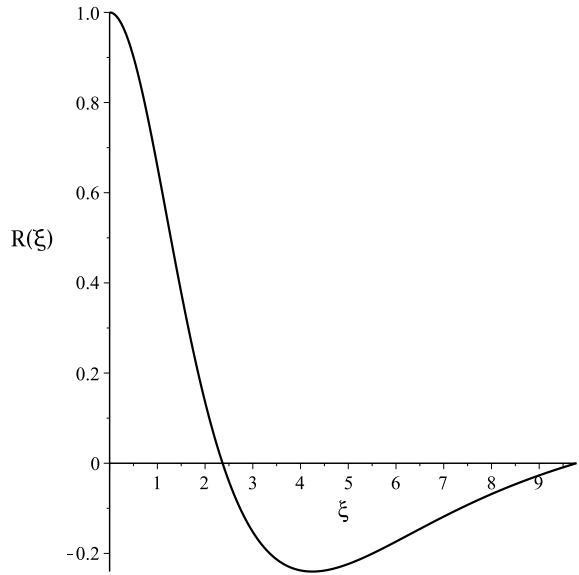} \caption{$R_{\xi}$ for $\omega=0.0167$ (a
solution of the energy condition).}%
\label{fig:fig4}%
\end{figure}

\begin{figure}[ptb]
\centering
\includegraphics[scale=0.4]{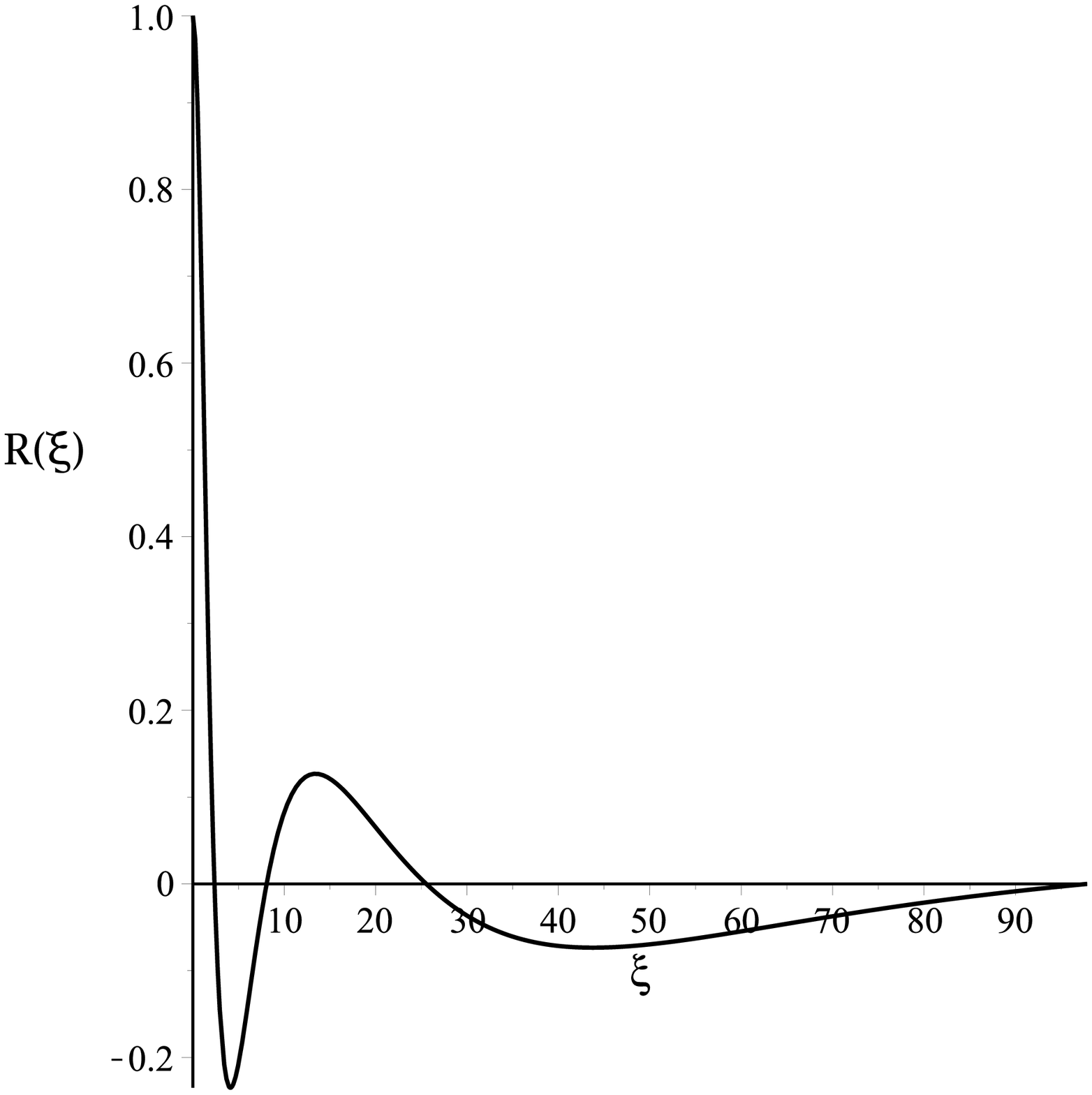} \caption{$R_{\xi}$ for $\omega=0.000167$
(a solution of the energy condition).}%
\label{fig:fig5}%
\end{figure}

\begin{figure}[ptb]
\centering
\includegraphics[scale=0.4]{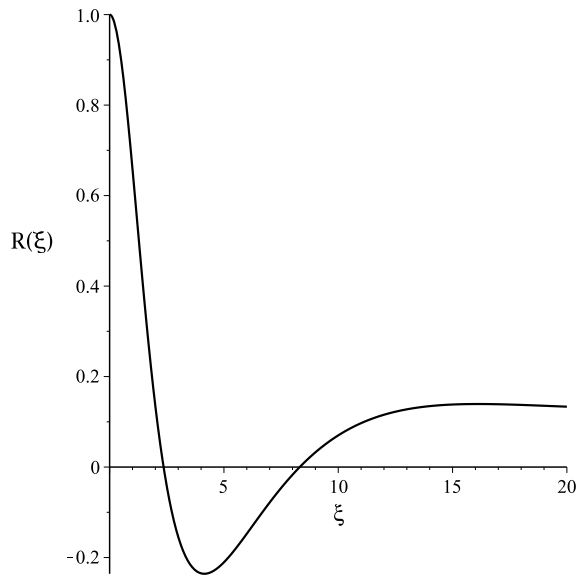} \caption{$R_{\xi}$ for $\omega=0.004$
(not a solution of the energy condition).}%
\label{fig:fig6}%
\end{figure}

\section{Summary and conclusion}

We studied the singular inverse square potential in the framework of quantum
mechanics with a GUP, which implies the existence of a minimal length. The
corresponding deformed Schr\"{o}dinger equation was established in coordinate
space by using the representation of Ref. \cite{brau}. By following Ref.
\cite{ha}, we transformed this equation into a second order differential
equation. We explicitly illustrated the regularizing effect that the
fundamental length plays on the singularity of the problem. Then, we solved
this equation analytically in terms of the confluent Heun function. The
problem of bound states has been discussed in detail and the energy spectrum
was derived in a natural way from the condition of square integrability of the
wave function. The results obtained here are identical to that of Ref.
\cite{djamil1}, where the momentum representation has been used. This
situation contradicts the conclusion drawn in the study of the hydrogen atom
problem \cite{boua3}, where the two representations have led to different
results. However, it is consistent with what was concluded in the study of the
harmonic oscillator \cite{chang}.

\begin{acknowledgments}
We thank Prof. Mahmut Horta\c{c}su for fruitful discussions and Prof. Michel
Bawin for reading the manuscript. We also thank the referee for interesting
comments and constructive criticisms that allowed us to improve the
manuscript. The work of DB is supported by the Algerian Ministry of Higher
Education and Scientific Research, under the CNEPRU Project No. D01720140007.
The research of TB is supported by TUBITAK, the Scientific and Technological
Council of Turkey.
\end{acknowledgments}

\end{document}